\documentclass[manuscript]{aastex} 
\usepackage{amsfonts,amsmath,graphicx,color}
 
\shorttitle{Notes about the distribution}
\shortauthors{Stephanovich, God{\l}owski}
 
\begin{document}
 
\title{The distribution of galaxies gravitational field stemming from their tidal interaction}

\author{Vladimir Stephanovich}
 
\affil{Uniwersytet Opolski, Institute of Physics, ul.  Oleska  48,
45-052 Opole, Poland}
\email{stef@uni.opole.pl}
 
\author{W{\l}odzimierz God{\l}owski}
 
\affil{Uniwersytet Opolski, Institute of Physics, ul.  Oleska  48,
45-052 Opole, Poland}
\email{godlowski@uni.opole.pl}
 
\begin{abstract}
We calculate the distribution function of astronomical objects (like galaxies and/or 
smooth halos of different kinds) gravitational fields due to their tidal interaction. For that 
we  apply the statistical method of \citet{chan43}, used there to calculate famous Holtzmark 
distribution. We show that in our approach the distribution function is never Gaussian, its form 
being dictated by the potential of interaction between objects. This calculation 
permits us to perform a theoretical  analysis of the relation between angular momentum and mass 
(richness) of the galaxy clusters.  To do so, we follow the idea of \citet{cat96} and \citet{heav88}. 
The main difference is that here we reduce the problem to discrete many-body case,
where all physical properties of the system are determined by the interaction potential $V({\bf r}_{ij})$.
The essence of reduction is that we use the multipole (up to quadrupole here) expansion 
of Newtonian potential so that all hydrodynamic, "extended" characteristics of an object like its 
density mass are "integrated out" giving its "point-like" characteristics like mass and quadrupole 
moment. In that sense we make no difference between galaxies and smooth components like halos. 
We compare our theoretical results with observational data.
\end{abstract}
 
\section{Introduction}

The problem of galaxies and their structures formation is one of the objectives of modern extragalactic 
astronomy and cosmology. There are many scenarios of structure formation
\citep{peb69,Sunyaew72,Zeldovich70,Doroshkevich73,Shandarin74,Dekel85,Wesson82,Silk83}, which are still 
important. This is because the new scenarios are essentially the modifications of
the old ones and can be classified according to the classical ones. Revised and improved structure formation 
scenarios can be found in various papers 
\citep{Lee00,Lee01,Lee02,Nav04,MO05,Bower05,Trujillio06,Brook08,vera11,Paz08,Shandarin12,Codis12,Varela12,Giah14}.
The crucial goal is to discriminate among different models of galaxies and their structures formation. 
The main controversy here is how galaxies acquire their angular moments, which yield subsequently the 
moments of galaxy clusters.

Presently the commonly accepted is spatially flat, homogeneous and isotropic $\Lambda$CDM model of the  
Universe. In such model, the structures were formed from the primordial adiabatic, nearly scale invariant 
Gaussian random fluctuations \citep{Silk68,Peebles70,Sunyaew70}. The most popular galaxy formation scenario,
 so-called hierarchic clustering model \citep{dor70,Dekel85,peb69} is based on this assumption.
The numerical simulations \citep {Bond96,Springel05,w1,w2} confirm that such mechanism could be realized 
in the Universe. In this mechanism, the large scale structure can appear from bottom to up as a consequence 
of gravitational interactions between galaxies. This means that galaxies are formed at the beginning with 
subsequent merger into larger clusters (structures). In this case, the galaxies spin angular momenta arise 
as a result of interaction with their neighbors. Original version of this model claims that the initial 
orientation of galaxy spins should be random. However, it had been shown later, that hierarchic clustering 
model admits naturally so-called Tidal Torque mechanism. In this mechanism, galaxies have their angular 
momenta aligned due to the coupling between the protogalaxy region and surrounding structure. The galaxy 
rotation in this mechanism is due to tidal interaction between galaxies \citep{Wesson82,White84} based 
on the ideas of \citet{Hoyle51}. The review of Tidal Torque scenario is presented by \citet{Sch09}. 
Note that while originally the Tidal Torque mechanism considers completely random distribution of spin 
angular momenta of galaxies, it has been shown later that the local tidal shear tensor can cause a local 
alignment of their rotational axes \citep{Dub92,cat96,Lee00,Lee01,Lee02,Nav04}. On the other hand, 
some authors like \citet{Brook08}, still argue that there is misalignment of angular momenta in the 
hierarchical clustering model. 

Different scenarios make different predictions concerning distribution of their angular momenta and especially 
about orientation of galaxies in structures (for review see \citet{g2011a}). Of course the final 
test of a given scenario is to compare its predictions with observations. It is possible to conclude that the 
observed variations in angular momentum represent simple but fundamental constraints for any model of galaxy 
formation \citep{Rom12,Joachimi15}.

The investigations of galaxies alignment show that it depends on the mass of the parent structure. Generally, 
the groups of galaxies and the small galaxy clusters reveal almost no alignment, while we observe such alignment 
 for rich galaxy clusters and superclusters. The alignment increases 
with mass of the structure \citep{Godlowski05,Aryal07,g10a,g2011a,g11c,g2011b}.  

Note that it is commonly agreed that there is no evidence for rotation of the groups and clusters of galaxies. 
 That implies that such structures do not rotate (for example  \citet{regos98,dia97,dia99,rines03,Hwang07}), 
see however \citet{kal05} for an opposite opinion. In this context, especially important is the result of
\citet{Hwang07} who examined the dispersions and velocity gradients in 899 Abell clusters and found
possible evidence for rotation in only six of them. Thus any non-zero angular momentum in groups and 
clusters of galaxies should arise only from possible alignment of galaxy spins and stronger alignment
mean larger angular momentum of such structures.

The aim of the present work is the theoretical  analysis of the influence of tidal interaction between 
objects like galaxies, their clusters as well as smooth halos
on their gravitational field distribution. For that we use the statistical method of \citet{chan43}.
To apply our result to observable quantities, we calculate the distribution function of the angular momenta. 
We do that for linear (corresponding to \citet{Zeldovich70} approximation in displacement field) and nonlinear 
regimes of fluctuation growth.
As in our method, the parameters of galaxy ensemble like their masses, radii and volumes enter problem as parameters, 
our calculation permits to trace possible relation between angular momentum and mass (richness) 
of the galaxy clusters. The above statistical method reveals the fact, that in the stellar systems, the derived 
distribution function cannot be Gaussian but rather belongs to the family of so-called "heavy-tailed distributions", 
see, e.g., \citet{kapkes92} for details. Moreover, choosing the cosmology on the base of corresponding 
Friedmann equation, our result permits to trace the time evolution of the distribution function of angular momenta 
and its mean value $L$. In our approach, we can also derive the well-known empirical relation between mean galaxy 
cluster moment $L$ and its mass $M$, $L \sim M^{5/3}$.

The paper is organized as follows. In Sections 2 and 3 we, based solely on the quadrupole (tidal) 
interaction potential between astronomical objects, calculate our universal (i.e. independent on the details of  
Lanrangian or Eulerian spaces) distribution function of gravitational fields $f(E)$. In the Section 4, on the 
base of the function $f(E)$, we calculate the distribution function of angular momenta $f(L)$ both in the linear 
and nonlinear Lagrangian approximation. We emphasize that as $f(E)$ does not depend on the Eulerian or Lagrangian 
picture, it can be used to calculate the distribution of any quantity (like momentum) of the astronomical objects 
(not only galaxy clusters but smooth component like halos as well) in any (linear or nonlinear) regime of fluctuation 
growth. 

\section{General formalism}

Unfortunately, the spin angular momentum is known only for very few galaxies and structures. Therefore, 
instead of the angular momentum by itself, the orientation of galaxies in each cluster is usually studied. 
This is the reason that here we are interested primarily in the absolute value of galaxies angular momenta.

We represent a matter (both luminous and dark) as the Newtonian self-gravitating fluid 
embedded in the Universe obeying corresponding Friedmann equation.
To obtain the tidal (i.e. shape distorting) interaction between the astronomical objects, we, similar to 
\citep{poisson98}, perform the multipole expansion of the Newtonian interaction potential between fluid 
elements. Truncating this expansion on the quadrupole terms, in the spirit of the article \citep{poisson98}, 
we write the Hamiltonian (total classical energy) of interaction between galaxies in the form 
\begin{eqnarray}
&&{\cal H}=- G\sum_{ij}Q_im_jV({\bf r}_{ij})\equiv-GM^2\sum_{ij}p_im_im_jV({\bf r}_{ij}), \label{star1} \\
&&V({\bf r}_{ij})=\frac 12 \frac{3\cos^2\theta_{ij} - 1}{r_{ij}^3},\nonumber 
\end{eqnarray}
where $G$ is the gravitational constant, $M=\sum_{i=1}^N m_i$ is the total mass of the ensemble, $p_i=Q_i/(m_iM^2)$, 
$Q_i$ is the quadrupole moment of the $i$-th galaxy, $m_i$ is its mass, $r_{ij}\equiv $ $|{\bf r}_{ij}|$, ${\bf r}_{ij}=$
 ${\bf r}_{j}-$ ${\bf r}_{i}$ is a relative separation between centres of galaxies and $N$ is their number.
The expression for $Q_i$ has the form \citep{poisson98}
\begin{equation}\label{star2}
Q_i=\int_{V_i} \rho_i({\bf x})|{\bf x}|^2P_2({\bf s}\cdot{\bf x})d^3x,\ P_2(x)=\frac 12 (3x^2-1),
\end{equation}
where $V_i$ is a volume of $i$-th galaxy, $\rho_i({\bf x})$ is a density of its mass.

Note, that expression \eqref{star1} generalizes the two-particle result of Ref. \citep{poisson98} 
on the ensemble of $N$ objects, splitting the interactions between them in pairwise manner. Such splitting 
is customary in condensed matter physics where the interacting many-body ensemble is represented by the sum 
of all possible pair interactions between particles $i$ and $j$ (for example 123=12+13+23), see 
Ref. \citep{majlis2000} and references therein in the context of magnetic systems. The same procedure is
 also used in astronomy \citep{astr}.  

The physics of the interaction \eqref{star1} is the following. Under the influence of the interaction with other 
astronomical objects, the shape of a given $i$-th object changes, which inflicts the variation of 
its density field $\rho_i({\bf x})$, Eq. \eqref{star2}. As the objects are situated randomly and 
have random shapes, the mass $m_i$ and quadrupole moment 
$Q_i$ of an object (like galaxy, galaxy cluster or smooth component) will vary randomly. This 
generates the random variations of the gravity field ${\bf E}_{quad}$ from 
these quadrupoles. Latter field is a gradient of the potential \eqref{star1} (divided by mass $m$) and has the form
 \begin{equation}\label{star3}
 {\bf E}_{quad}({\bf r})={\bf i}_rE_0\frac{3\cos^2\theta - 1}{r^4},\ E_0=\frac{GQ}{2},
 \end{equation}
where ${\bf i}_r$ is the unit vector in the direction of the radius-vector.

As we cannot solve the random many-body problem \eqref{star1} exactly, different approaches had been used for its 
approximate solution. The simplest approach in condensed matter physics is so-called mean field approximation, where 
the fluctuating (electric, magnetic, elastic in condensed matter) field, acting on the specific particle is substituted 
by some ensemble averaged, mean field, see, e.g. \citep{majlis2000}. This approach does not take into account the 
particle clustering (i.e. two-, three-, four-particle cluster in the field of the rest of likewise clusters) and is not 
suitable to describe the systems with disorder, which is also our case. In the disordered case the adequate approach is
 statistical method similar to that applied by Chandrasekhar \citep{chan43} to describe the fluctuations of the force 
acting on a specific star from the rest of stellar system. In this method, the distribution function of random
 gravitational field is introduced so that any observable quantity of an ensemble (like orbital moment, energy etc) can 
be expressed by averaging the corresponding single-particle quantity with the above distribution 
function, see Refs. \citep{stef97},\citep{semstef02}, \citep{semstef03} and references therein, where the statistical 
method had been applied to disordered solids.   

\section{Distribution function of quadrupolar fields}

According to statistical method, very similar to that from Chandrasekhar \citep{chan43}, the distribution function of
 random quadrupolar fields reads (see \citet{stef97, semstef03} and references therein)
\begin{equation}\label{star8}
f({\bf E})=\overline{\delta({\bf E}-{\bf E}_i)},
\end{equation}
where $\delta(x)$ is Dirac $\delta$ - function, ${\bf E}_i$ is given by the expression \eqref{star3} and bar means the 
averaging over spatial (and any other possible) disorder. We note that the above statistical method has frequently been 
used to describe the physical properties of disordered solids like dielectrics \citep{stef97} and/or magnetic 
semiconductors \citep{semstef03}.

The physical meaning of distribution function \eqref{star8} is as follows. If we do not have any randomness in the 
galaxies ensemble (for instance, all galaxies are similar) the distribution function is just delta-peak, centered in
 the corresponding field ${\bf E}_i$. If we have a disorder in a system, the averaging broadens this delta-peak, giving 
rise to "bell-shaped" continuous probability distribution. During this averaging the index $i$ will be "swallowed" as 
we average over the stellar ensemble.

To perform the prescribed averagings, we pass to the integral representation of Dirac $\delta$ - function to obtain
\begin{equation}\label{star9}
f({\bf E})=\overline{\delta({\bf E}-{\bf E}_i)}=\frac{1}{(2\pi)^3}\int_{-\infty}^\infty e^{i{\bf E}{\boldsymbol \rho}}\overline{e^{-i{\boldsymbol \rho}{\bf E}_i}}d\rho_x d\rho_y d\rho_z.
\end{equation} 
 
The explicit averaging in eq. \eqref{star9} is based on the fact that the mass and quadrupole moment of the galaxies 
in the volume $V$ obey the uniform distribution with probability density equal to $1/V$. This is equivalent to the fact
that the fluctuations of above galaxies parameters obey Poisson distribution \citep{chan43}, see also \citep{zol86} for 
purely mathematical treatment of this question. In this case the averaging for single galaxy yields
\begin{equation}\label{star9a}
[\overline{e^{-i{\boldsymbol \rho}{\bf E}_i}}]_1=\frac{1}{4\pi V}\int_Ve^{-i{\boldsymbol \rho}{\bf E}({\bf r})}d^3r,
\end{equation}
while for the $N$ galaxies ensemble the corresponding average equals to 
$\overline{e^{-i{\boldsymbol \rho}{\bf E}_i}}\equiv$ $[\overline{e^{-i{\boldsymbol \rho}{\bf E}_i}}]_1^N$, i.e. the 
single-galaxy average, raised to power $N$ \citep{chan43}. 

We note here that the above averaging procedure implies that the astronomical objects are similar
 to each other. However, it can be shown (see \citet{semstef02} and references therein), that such approach takes 
into account the pair clusters of galaxies or other spatially disordered constituents in the systems other then stellar.
 The higher order clusters (three, four etc objects) can be taken into account along the lines of
 Ref. \citep{ziman} which would require to solve the chain of kinetic equations for $n=3,4$ etc bodies distribution 
functions. On the other hand, since the above procedure accounts exactly for pair clusters, the many-body clusters
 can be considered by the splitting them into corresponding pairs. This means that our approach takes also the 
many-particle clustering into account. However, in the future, we are going to elaborate the above $n$ - bodies 
procedure and compare it to our present results.  

We have finally
\begin{eqnarray}
&&\overline{e^{-i{\boldsymbol \rho}{\bf E}_i}}=\lim_{N\to\infty, V\to \infty}\left[\frac{1}{4\pi V}\int_Ve^{-i{\boldsymbol \rho}{\bf E}({\bf r})}d^3r\right]^N
\equiv \lim_{N\to\infty, V\to \infty}\left[\frac 1V \int_V U({\bf r})d^3r\right]^N\equiv \nonumber \\
&&\equiv \lim_{N\to\infty, V\to \infty} \left\{1-\frac 1V\int_V \biggl[1-U({\bf r})\biggr]d^3r\right\}^N= 
\lim_{N\to\infty}  \left\{1-\frac nN\int_V \biggl[1-U({\bf r})\biggr]d^3r\right\}^N\equiv \nonumber \\
&&\equiv \exp\biggl[-n\int_V \bigl[1-U({\bf r})\bigr]d^3r\biggr],\nonumber \\
&&U({\bf r})\equiv \frac{1}{4\pi}\int e^{-i\rho E\cos \theta}
\sin \theta d\theta d\varphi=\frac{\sin \rho E({\bf r})}{\rho E({\bf r})},
\label{star10}
\end{eqnarray}
where $E({\bf r})$ is given by Eq. \eqref{star3}. 

The expression for $U({\bf r})$ signifies the averaging over random angle between vectors $\rho$ and ${\bf E}$. Also, 
we denote in \eqref{star10} $N$ to be number of galaxies so that $n=N/V=const$ is a density of galaxies in a given 
volume. 

Combining the equations \eqref{star9} and \eqref{star10}, we obtain following expression for the distribution function
\begin{equation}\label{star12}
f({\bf E})=\frac{1}{(2\pi)^3}\int_{-\infty}^\infty\ e^{i{\bf E}{\boldsymbol \rho}-F(\rho)}d^3\rho,\  F(\rho)=
n\int_V\left[1-\frac{\sin \rho E({\bf r})}{\rho E({\bf r})}\right]d^3r.
\end{equation}
We see that $F(\rho)$ is indeed the characteristic function for random gravitational fields distribution. Note that 
the galaxy clustering can be better considered if we assume that the density of galaxies is 
not a constant but rather $n=n({\bf r})$. Then, similar to the case of disordered magnetic semiconductors 
[see Eq. (6b) of \citet{semstef03}], the function $n({\bf r})$ should be put under the integral sign in \eqref{star12} 
to give
\begin{equation}\label{sta12}
F(\rho)=
\int_Vn({\bf r})\left[1-\frac{\sin \rho E({\bf r})}{\rho E({\bf r})}\right]d^3r.
\end{equation}
If we specify the empirical dependence $n({\bf r})$, the characterstic function $F(\rho)$ can be calculated only 
numerically. 

Below we shall calculate this function without any assumptions analytically for the simpler case $n=$ 
const. One more way to consider the effects of clustering is to account for inhomogeneous distribution of masses 
(and/or quadrupolar moments) in the ensemble. This can be done along the lines of Ref. \citep{chan43}, where the 
distribution function of masses $\tau(m)$ had been introduced. In our case, however, the situation will be not 
that simple as we are dealing with more complicated object like quadrupole moment of an object. 
In the present publication we do not consider this effect especially in view that there is large ambiguity in 
determination of function $\tau(m)$ from the astronomical observation data. However, we are going to incorporate
the dependence $\tau(m)$ in our consideration in future.

We note here that distribution function $f({\bf E})$ \eqref{star12} is by no means Gaussian. We will show  that function
 \eqref{star12} does not admit Gaussian limit so that for problem of interaction between gravitational quadrupoles (and
 actually higher order terms like octupoles etc) the distribution function of gravitational fields and angular momenta 
is never Gaussian. To demonstrate that, we note that it had been
shown earlier for condensed matter systems (ferroelectrics in \cite{stef97} and magnetic semiconductors in 
\cite{semstef03} ) that Gaussian limit corresponds to large density $n\to \infty$ of electric dipoles (ferroelectrics) 
or spins (magnetic systems). 

It had also been shown by \cite{stef97} and \cite{semstef03} that limit $n\to \infty$ corresponds to small 
Fourier variable $\rho$ in the equation \eqref{star12}. This means that to obtain the Gaussian limit of the 
distribution function \eqref{star12}, we should expand its characteristic function $F(\rho)$ in small $\rho$. In first 
nonvanishing (Gaussian) approximation in small $\rho$ this procedure yields
 \begin{equation}\label{star12a}
 F_G(\rho)=\frac{n\rho^2}{6}\int_V E^2({\bf r})d^3r
 \end{equation}
To derive equation \eqref{star12a}, we use the expansion $\sin x/x\approx 1-x^2/6 + ...$, valid at $x \to 0$.
The explicit rewriting of \eqref{star12a} with respect to \eqref{star3} yields
 \begin{equation}\label{star12b}
  F_G(\rho)=\frac 13 \pi n\rho^2 \int_0^\pi (3\cos^2\theta -1)^2\sin \theta d\theta\ \int_0^\infty \frac{dr}{r^6},
 \end{equation}
where the last integral over $r$ is divergent at small $r$. We note here that in the solids the Gaussian limit exists 
(i.e. the integral \eqref{star12b} becomes convergent) due to presence of short range terms $\sim \exp(-r/r_c)$ 
($r_c$ is so-called correlation radius, defining the range of interaction) in the interaction potential between dipoles 
or spins. As the exchange interaction between spins is always of short range, this feature is peculiar to magnetic 
systems, see \cite{semstef03} and references therein. This means that due to long-range quadrupole-quadrupole 
interaction between galaxies, $\sim r^{-4}$, the distribution function of their fields and/or orbital moments (see
below) is never Gaussian.

The explicit expression for $F(\rho)$ reads
\begin{equation}\label{star13}
F(\rho)=n\int_0^\infty r^2dr\int_0^\pi \sin\theta d\theta \int_0^{2\pi}d \varphi\ \left[1-\frac{\sin\frac{\rho E_0(3\cos^2\theta-1)}{r^4}}{\frac{\rho E_0(3\cos^2\theta-1)}{r^4}}\right].
\end{equation}
We first perform the integration over $r$ in \eqref{star13}. For that we denote $b/r^4=x$ ($b=\rho E_0(3\cos^2\theta-1)$)
 to obtain 
\begin{equation}\label{star14}
K=\int_0^\infty r^2dr \left(1-\frac{\sin Q}{Q}\right)=\frac 14 |b|^{3/4}\int_0^\infty \frac{dx}{x^{7/4}}\left(1-\frac{\sin x}{x}\right).
\end{equation}
The last integral can be calculated analytically to give
\[
\int_0^\infty \frac{dx}{x^{7/4}}\left(1-\frac{\sin x}{x}\right)=\frac{8}{21}\sqrt{2-\sqrt{2}}\  \Gamma\left(\frac 14\right)\approx
1.05711302,
\]
where $\Gamma(x)$ is $\Gamma$ - function \citep{abr}.
This finally yields
\begin{equation}\label{star15}
K=0.264278 |b|^{3/4} \equiv A |b|^{3/4}.
\end{equation}
With respect to \eqref{star15} we have 
\begin{equation}\label{star16}
F(\rho)=2\pi nA(\rho E_0)^{3/4}\int_0^\pi \sin \theta d\theta |3\cos^2\theta-1|^{3/4}.
\end{equation}
The auxiliary integral $K_1$ can be calculated numerically to give
\begin{equation}\label{star17}
K_1=\int_{-1}^1dx|3x^2-1|^{3/4}=2\int_{0}^1dx(3x^2-1)^{3/4}=1.581940777.
\end{equation}
With respect to \eqref{star17}, $F(\rho)$ assumes the form
\begin{equation}\label{star18}
F(\rho)= 2\pi n\cdot 0.41807255\cdot (\rho E_0)^{3/4}.
\end{equation}

Substitution of \eqref{star18} into \eqref{star12} gives
\begin{eqnarray}
&&f(E)=\frac{1}{2\pi^2E}\int_0^\infty \rho \ e^{-\alpha \rho^{3/4}}\ \sin \rho E\
d\rho,\label{star19}\\
&& \alpha= 2\pi n \cdot 0.41807255\cdot E_0^{3/4}. \nonumber
\end{eqnarray}
The expression \eqref{star19} is the main theoretical result of the present paper, constituting the final answer for 
distribution function of gravitational fields moduli. It is seen that distribution function \eqref{star19} depends 
parametrically on the galaxies density $n$, as well as on average galaxy quadrupole moment. To the best of our 
knowledge, neither distribution function \eqref{star19} nor its explicit dependence on $n$ and $Q$ has been known 
previously. The integral \eqref{star19} can be calculated only numerically.

The normalization condition for distribution function \eqref{star19} looks like
\begin{equation}\label{sta1}
4\pi \int_0^\infty E^2f(E)dE=1.
\end{equation}

We check explicitly
\[
I=4\pi \int_0^\infty E^2f(E)dE=\frac{2}{\pi}\int_0^\infty E\sin \rho EdE \int_0^\infty
\rho\ d\rho\ e^{-\alpha \rho^{3/4}}.
\]
With respect to the relation $\int_0^\infty E\sin \rho EdE=-\pi \delta'(\rho)$, the integral $I$ can be calculated
\begin{equation}\label{sta2}
I=-2 \int_0^\infty \rho\ \delta'(\rho) d\rho\ e^{-\alpha \rho^{3/4}}
=\frac{d}{d\rho}\left[\rho\ e^{-\alpha \rho^{3/4}}\right]_{\rho=0}=1.
\end{equation}

To derive the result \eqref{sta2}, we use the following identity for $n$ -th derivative of Dirac $\delta$ function
\[
\int_{-\infty}^\infty f(x) \delta^{(n)}(x)dx=-\int_{-\infty}^\infty \frac{df}{dx} \delta^{(n-1)}(x)dx.
\]

We finally mention the difference of our approach to the problem of angular moments distribution and that of
 Refs. \citep{cat96} and \citep{heav88}. The main difference is that we consider the discrete 
many-body problem, stemming from multipole expansion (up to quadrupole here) of Newtonian interaction 
potential between fluid elements. In such approach, all physical 
properties of the system are determined by the interaction potential $V({\bf r}_{ij})$ \eqref{star1}. 
In the case of disorder, the distribution function of random gravitational fields is also completely 
determined by the form of potential $V({\bf r}_{ij})$. To derive the distribution function 
of random gravitational fields, we use the statistical method of \citet{chan43}. As we 
have shown above, within this method, the distribution function is never Gaussian for any long-range 
potential, obtained as multipole expansion of Newtonian one. At the same time, all previous approaches 
postulated (rather then derived) the distribution function in Gaussian form, which, in our opinion, 
does not reflect the physical nature of long-range gravitational multipole interaction, 
which generates distribution functions with long tails. 

 \subsection{Numerical calculation of $f(E)$. Dimensionless variables.}

Following \citet{chan43} for the case of Holtzmark function, we introduce the dimensionless variables 
$\rho E=x$ and $\beta= E/\alpha^{4/3}$. In these variables the integral \eqref{star19} renders to
\begin{eqnarray}
&&f(\beta)=\frac{1}{2\pi^2\beta^3\alpha^4}\int_0^\infty x\ \sin x \ \exp{\left[-\left(\frac x\beta\right)^{3/4}\right]}dx\equiv
\frac{H(\beta)}{4\pi\beta^2\alpha^4},\label{star20a} \\
&&H(\beta)=\frac{2}{\pi \beta}\int_0^\infty x\ \sin x \ \exp{\left[-\left(\frac x\beta\right)^{3/4}\right]}dx. \label{star20b}
\end{eqnarray}
The physical meaning of the function $H(\beta)$ is that as it proportional to $\beta^2 f(\beta)$, it 
is just the integrand in Eq. \eqref{sta1}, being the effective one-dimensional distribution function of 
random fields.  In other words, the normalization condition for $H(\beta)$ assumes one-dimensional form
\begin{equation}\label{st1a}
\int_0^\infty H(\beta)d\beta=1.
\end{equation}
In this case, the average value of dimensionless random field $\beta$ reads
\begin{equation}\label{st1b}
{\bar \beta}=\int_0^\infty \beta H(\beta)d\beta.
\end{equation}
Function $H(\beta)$ will be calculated numerically below. 

\subsection{Asymptotics of distribution function $f(\beta)$}

We begin with asymptotics of $H(\beta)$. At $\beta \to 0$ we make substitution $x/\beta=t$ to obtain from \eqref{star20b}

\begin{equation}\label{star21}
H(\beta)=\frac{2\beta}{\pi}\int_0^\infty\ t\sin \beta t\ e^{-t^{3/4}}dt \approx \frac{2\beta}{\pi}\int_0^\infty\ t \left(\beta t-\frac 16 (\beta t)^3+...\right)\ e^{-t^{3/4}}dt.
\end{equation}
The first term of the expansion \eqref{star21} yields
\begin{equation}\label{star22}
H(\beta \to 0)= \frac{2\beta^2}{\pi}\int_0^\infty t^2e^{-t^{3/4}}dt=\frac{8\beta^2}{3\pi}\int_0^\infty z^3e^{-z}dz\equiv \frac{48}{3\pi}\beta^2.
\end{equation}

At $\beta \to \infty$ we expand the Eq. \eqref{star20b} over small parameter $1/\beta$. We obtain in the first
 approximation
\begin{eqnarray}
&&H(\beta \to \infty)=\frac{2}{\pi \beta}\lim_{\delta \to 0}\int_0^\infty\ x\sin x \left(1-\frac{x^{3/4}}{\beta^{3/4}}\right)e^{-\delta x}dx=\nonumber \\
&&=-\frac{2}{\pi \beta^{7/4}}\lim_{\delta \to 0}\int_0^\infty\ x^{7/4}\sin x e^{-\delta x}dx=\nonumber \\
&&=\frac{2}{\pi \beta^{7/4}}\Gamma(11/4)\cos \frac \pi 8 \approx 0.945972642 \beta^{-7/4}. \label{star23}
\end{eqnarray}
To derive expression \eqref{star23} we take into account that 
$\lim_{\delta \to 0}\int_0^\infty x\sin x e^{-\delta x}dx=0$.

Having asymptotics $H(\beta)$, we calculate those for $f(\beta)$ with the help of relation \eqref{star20a}: 
\begin{equation}\label{star24}
f(\beta)=\left\{\begin{array}{c}
\frac{4}{\pi^2\alpha^4},\hspace*{1.7cm} \beta \to 0 \\ \\ 
\frac{0.945972642}{4\pi \alpha^4}\ \beta^{-15/4},\ \beta \to \infty.
\end{array}\right.
\end{equation}
Here $\alpha$ is given by Eq. \eqref{star19}. The asymptotics \eqref{star24} shows that $f(\beta)$ does not depend 
on $\beta$ at small $\beta$ and decays at large $\beta$. This shows that although normalization integral is convergent 
(we recollect that normalization condition for $f(\beta)$ looks like $\int_0^\infty \beta^2 f(\beta)d\beta=1$), 
already first moment does not exist. This can also be seen from large $\beta$ asymptotics of $H(\beta)$ \eqref{star23}. 
The expression \eqref{star24} is a confirmation of the fact that function $f(\beta)$ belongs to the class
of heavy-tailed distributions.

\section{Distribution function of angular momenta}

Our aim is to calculate the distribution function of galaxies angular momenta ${\bf L}_i$. For that we need a relation 
between the angular momentum ${\bf L}$ of a galaxy and its gravitational field ${\bf E}_{quad}({\bf r})$ \eqref{star3}.
The expression for components $L_\alpha$ ($\alpha=x,y,z$) of ${\bf L}$ has been derived in the 
form of perturbation series in small Lagrangian coordinate ${\bf q}$. The first order terms are defined by 
Eq. (11) of Ref. \citep{cat96}, while second order ones by Eq.  (28) of the follow-up article \citep{ca96}.
Both expressions have identical structure and can be written in the form
 \begin{equation}\label{star4}
 L_\alpha^{(i)}=f_i(t)\varepsilon_{\alpha \beta \gamma}E_{i \beta \sigma}I_{\sigma \gamma},\ \alpha,\beta,\gamma,\sigma = x,y,z,
 \end{equation}
where index $i=1,2$ denotes the order of perturbation theory, $\varepsilon_{\alpha \beta \gamma}$ is
 Levi-Civita symbol, $E_{\beta \sigma}$ are components of quadrupole (tidal) field \eqref{star3} and 
$I_{\sigma \gamma}$ are components of inertia tensor. 
Accordingly, functions $f_1(t)=a^2(t)\dot D(t)$ and $f_2(t)=\dot E(t)$ (dot means time derivative) are 
known functions of time, calculated from the differential equations, derived in $i$ - th order of 
perturbation theory \citep{bouchet92}. The explicit form of these equations read 
\begin{eqnarray}
&&t_0^2\ddot D(t)+a(t)D(t)=0, \label{first} \\
&&t_0^2\ddot E(t)+a(t)E(t)=-a(t)D(t)^2, \label{sec}
\end{eqnarray}
where $0\leq t <\infty$ is dimensional physical time and $t_0$ is some characteristic time, depending 
of the cosmological model considered. We will choose this quantity below. The dimensionless function 
$a(t)$ (so-called scale factor) is determined from zero-order perturbative equation, which indeed is 
first Friedmann equation, depending again on chosen cosmological model. General form of this equation reads
\begin{equation}\label{frid}
\frac{H^2}{H_0^2}=\Omega_R a^{-4}+\Omega_M a^{-3}+\Omega_k a^{-2}+\Omega_\Lambda.
\end{equation}
Here $H=\dot{a}/a$ is Hubble parameter ($\dot a\equiv$ $da/dt$), $H_0$ is Hubble constant and 
$\Omega_i$ ($i=$ R,M,k,$\Lambda$) are corresponding density parameters taken at present time, when $a(t)=1$. 
Specifically, $\Omega_R$ is radiation density, $\Omega_M$ is matter (dark plus baryonic) density, $\Omega_k$ 
is co-called spatial curvature density and $\Omega_\Lambda$ is cosmological constant or vacuum density, 
$\Omega_\Lambda=\Lambda/(3H_0^2)$, where $\Lambda$ is cosmological and $H_0$ is Hubble constant. For our 
calculations of distribution function of angular momenta, we will choose flat
$\Lambda$CDM model of the Universe, keeping in the equation \eqref{frid} only $\Omega_M$ and 
$\Omega_\Lambda$ terms, $\Omega_M+\Omega_\Lambda=1$.

Although the components of the gravitational tidal (shear) field ${\bf{E}}$ are different 
in the first and second orders of perturbation theory, for our purposes it is sufficient to consider them 
to be the same as they are simply the arguments of distribution function \eqref{star19}. Denoting 
$b(t)=$ $f_i(t)$ and omitting index $i$ in the components of tidal field ${\bf{E}}$, we can rewrite 
the relation \eqref{star4} explicitly:
\begin{eqnarray}
L_x/b(t)=E_{yx}I_{xz}+E_{yy}I_{yz}+E_{yz}I_{zz}-E_{zx}I_{xy}-E_{zy}I_{yy}-E_{zz}I_{zy},\nonumber \\
L_y/b(t)=E_{zx}I_{xx}+E_{zy}I_{yx}+E_{zz}I_{zx}-E_{xx}I_{xz}-E_{xy}I_{yz}-E_{xz}I_{zz},\nonumber \\
L_z/b(t)=E_{xx}I_{xy}+E_{xy}I_{yy}+E_{xz}I_{zy}-E_{yx}I_{xx}-E_{yy}I_{yx}-E_{yz}I_{zx}.\label{star6}
\end{eqnarray}
Taking into account the symmetry relations $I_{ab}=I_{ba}$ and $E_{ab}=E_{ba}$ and leaving 
only $E_{zz}$, we obtain from \eqref{star6}
\begin{eqnarray}
&&L_x=-b(t)E_{zz}I_{yz},\ L_y=b(t)E_{zz}I_{xz},\ L_z=0. \nonumber \\ 
&&L=\sqrt{L_x^2+L_y^2+L_z^2}=b(t)E_{zz}\sqrt{I_{xz}^2+I_{yz}^2}\equiv L_0E,\label{star7} \\
&&L_0=L_0(t)=f_i(t)\sqrt{I_{xz}^2+I_{yz}^2}. \nonumber
\end{eqnarray}
The expression \eqref{star7} constitutes linear relation between angular momentum and tidal field moduli
 both in linear ($i=$1) and nonlinear ($i=$2) regimes.

As the relation between gravitational field modulus and angular moment \eqref{star7} is linear, the shape of 
distribution function of angular moments $f(L)$ is similar to that of gravitational fields. The explicit transition 
from $f(E)$ \eqref{star19} to $f(L)$ can be accomplished combining the expression \eqref{star7} and known relation 
from the theory of probability
\begin{equation}\label{star25}
f(L)=f[E(L)]\left|\frac{dE}{dL}\right|,
\end{equation}
which yields
\begin{equation}\label{star26}
f(L)=\frac{1}{2\pi^2L}\int_0^\infty \rho \ e^{-\alpha \rho^{3/4}}\ \sin \left(\rho\frac{L}{L_0(t)}\right) \
d\rho,
\end{equation}
where $L_0(t)$ is given by expression \eqref{star7}. Passing to dimensionless variables 
\begin{equation}\label{star26a}
\rho \frac{L}{L_0}=x,\ \lambda=\frac{L}{L_0\alpha^{4/3}}
\end{equation}
generates the following pair of functions similar to the case of gravitational fields distribution
\eqref{star20a}, \eqref{star20b} 

\begin{eqnarray}
&&f(\lambda)=\frac{1}{2\pi^2\lambda^3\alpha^4L_0}\int_0^\infty x\ \sin x \ \exp{\left[-\left(\frac x\lambda\right)^{3/4}\right]}dx\equiv
\frac{H(\lambda)}{4\pi\lambda^2\alpha^4L_0},\label{star27a} \\ \nonumber \\
&&H(\lambda)=\frac{2}{\pi \lambda}\int_0^\infty x\ \sin x \ \exp{\left[-\left(\frac x\lambda\right)^{3/4}\right]}dx. \label{star27b}
\end{eqnarray}
In this case the effective 1D distribution function $H(\lambda)$ is similar to that from the gravitational fields 
\eqref{star20b} and we have the asymptotics \eqref{star24} (divided by $L_0$) for the distribution function of momenta.

\begin{figure}[tbh]
\centering
\includegraphics[width=0.9\columnwidth]{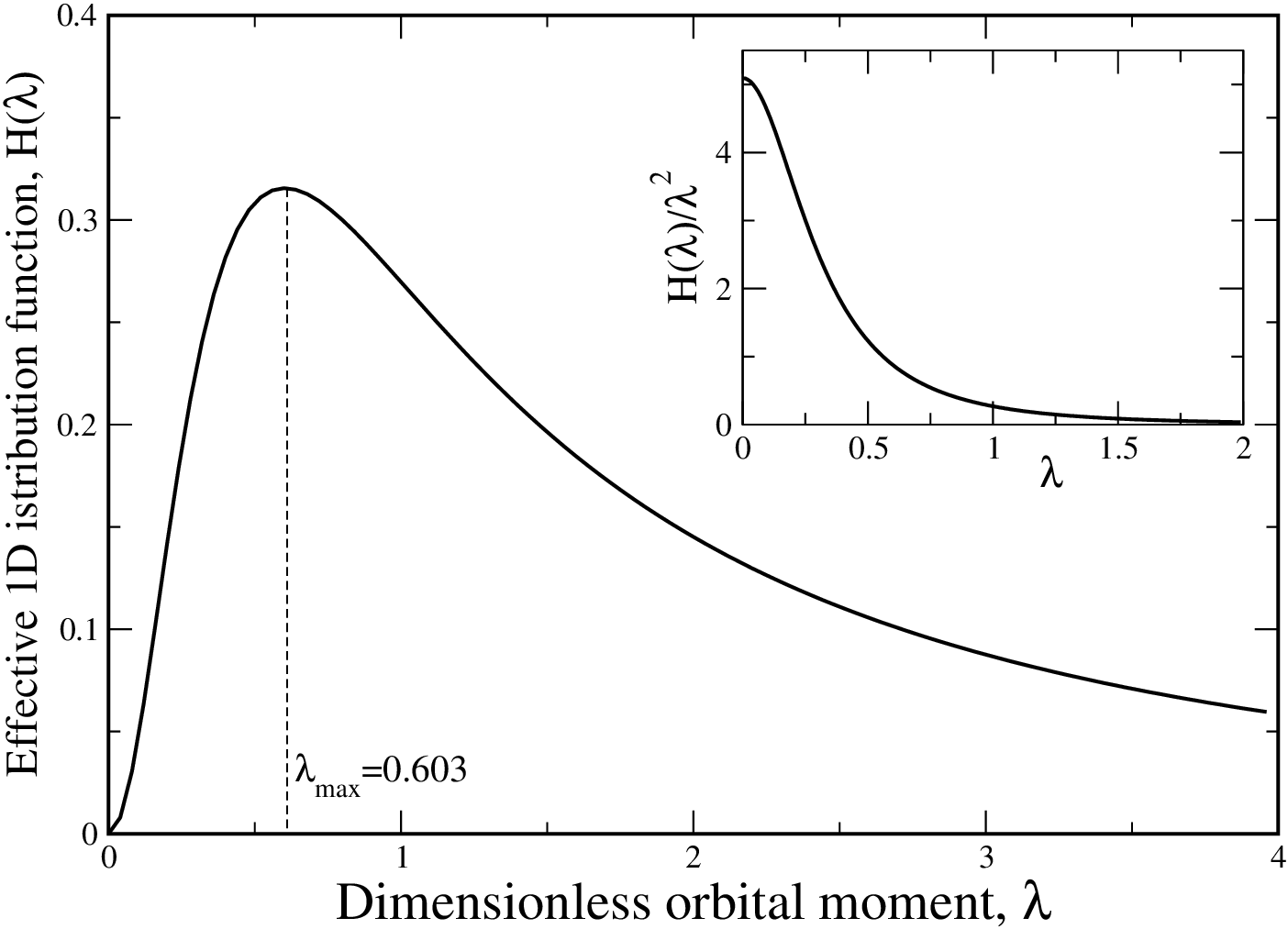}
\caption{The effective 1D distribution function  \eqref{star27b} of dimensionless orbital moments $H(\lambda)$. 
The distribution function of dimensionless gravitational fields $H(\beta)$ \eqref{star20b} has the same shape. 
Dashed line shows the argument $\lambda_{max}$ \eqref{star28}, corresponding to maximum of $H(\lambda)$. Inset 
depicts real 3D distribution function $4\pi \alpha^4 L_0f(\lambda)=$ $H(\lambda)/\lambda^2$.} \label{fig1}
\end{figure}

The effective 1D distribution of fields or momenta is shown in Fig. \ref{fig1}. It is seen that 
this function has characteristic bell shape and is asymmetric. Asymptotics \eqref{star23} shows that the 
integral, defining the mean value of the galaxy orbital moment \eqref{st1b}, is divergent.
To estimate the most probable value of the orbital moment, we calculate $\lambda_{max}$, corresponding to the maximum of 
distribution function $H(\lambda)$, see Fig. \ref{fig1} for details. This 
situation is typical for so-called heavy-tailed distributions like Cauchy one $f(x)\sim $ $(a^2+x^2)^{-1}$. 
Such distributions can be met very frequently in all branches of physics, dealing with random processes 
and ranging from condensed matter physics to chemical kinetics and econophysics, see, e.g. \citet{kapkes92}.  
As Cauchy and many other heavy-tailed distributions do not admit a first moment (corresponding integral 
is divergent), the maximum of such probability density function is usually taken as a measure of its mean value. 

The analysis of this mean value will permit us to derive some useful relations, which 
earlier had been taken as empirical ones. To perform this analysis we adopt the simplest possible CDM 
cosmology in the first order of perturbation theory, where $a(t)=D(t)=(t/t_0)^{2/3}$ \citep{dor70} 
so that $L_0=\frac{2I}{3t_0}\tau$, $\tau=t/t_0$, $I=\sqrt{I_{xz}^2+I_{yz}^2}$. 
The solution of the equation $dH/d\lambda=0$ reads
\begin{equation}\label{star28}
\lambda_{max}=0.602730263.
\end{equation}
In dimensional units \eqref{star26a} we have from \eqref{star28}
\begin{eqnarray}
&&L_{max}=0.602730263\cdot(2\pi\cdot 0.41807255)^{4/3}n^{4/3}L_0\frac{GQ}{2}=\nonumber \\
&&=1.09228264 \left(\frac NV\right)^{4/3}L_0GQ,\label{lm1} \\
&&L_{max}=0.7281884\left(\frac NV\right)^{4/3}\frac{t}{t_0^2}GIQ \approx \kappa\left(\frac NV\right)^{4/3}\frac{t}{t_0^2}GR^4m^2, \label{lm2}
\end{eqnarray}
where $\kappa$ is a constant of order unity. To derive the last equation \eqref{lm2}, we estimate (on the base of
 Eq. \eqref{star2}) both galaxy quadrupole moment $Q$ and its mean inertia moment $I$ as being proportional to $mR^2$, 
where $m$ is galaxy mass and $R$ is its mean radius. If we estimate volume $V$ as $V=R^3$, we see that $R$ cancels down 
in Eq. \eqref{lm2} so that we obtain $L_{max} \sim (t/t_0^2) m^2 N^{4/3}$. 

On the other hand, we can suppose that volume $V$ is related to the galaxy cluster so that its value $V=R_A^3$, 
where $R_A$ is a mean radius of the cluster, which is proportional to the autocorrelation radius 
\citep{autocor,autocor1,astr,lin,tucker,long08}.
Although $n$  is still a constant for any particular cluster, it varies from cluster to cluster with 
increasing richness $N$.
In this case we may rewrite $N=M/m$ to obtain the different form of expression for $L_{max}$
\begin{equation}\label{lm3}
L_{max} \sim \frac{t}{t_0^2}\ \left(\frac{R}{R_A}\right)^4\ m^{2/3}M^{4/3},
\end{equation}
which does not contain $\rho$.
It seems formally, that Eq. \eqref{lm3} implies that $L_{max} \sim t$, but the time evolution of galaxy 
radius $R$ and mean cluster radius $R_A$, being very complex astrophysical process \citep{long08}, can 
complicate real time dependence $L_{max}(t)$ a lot. This question needs to be studied additionally.

\begin{figure}
\centering
\includegraphics[width=0.9\columnwidth]{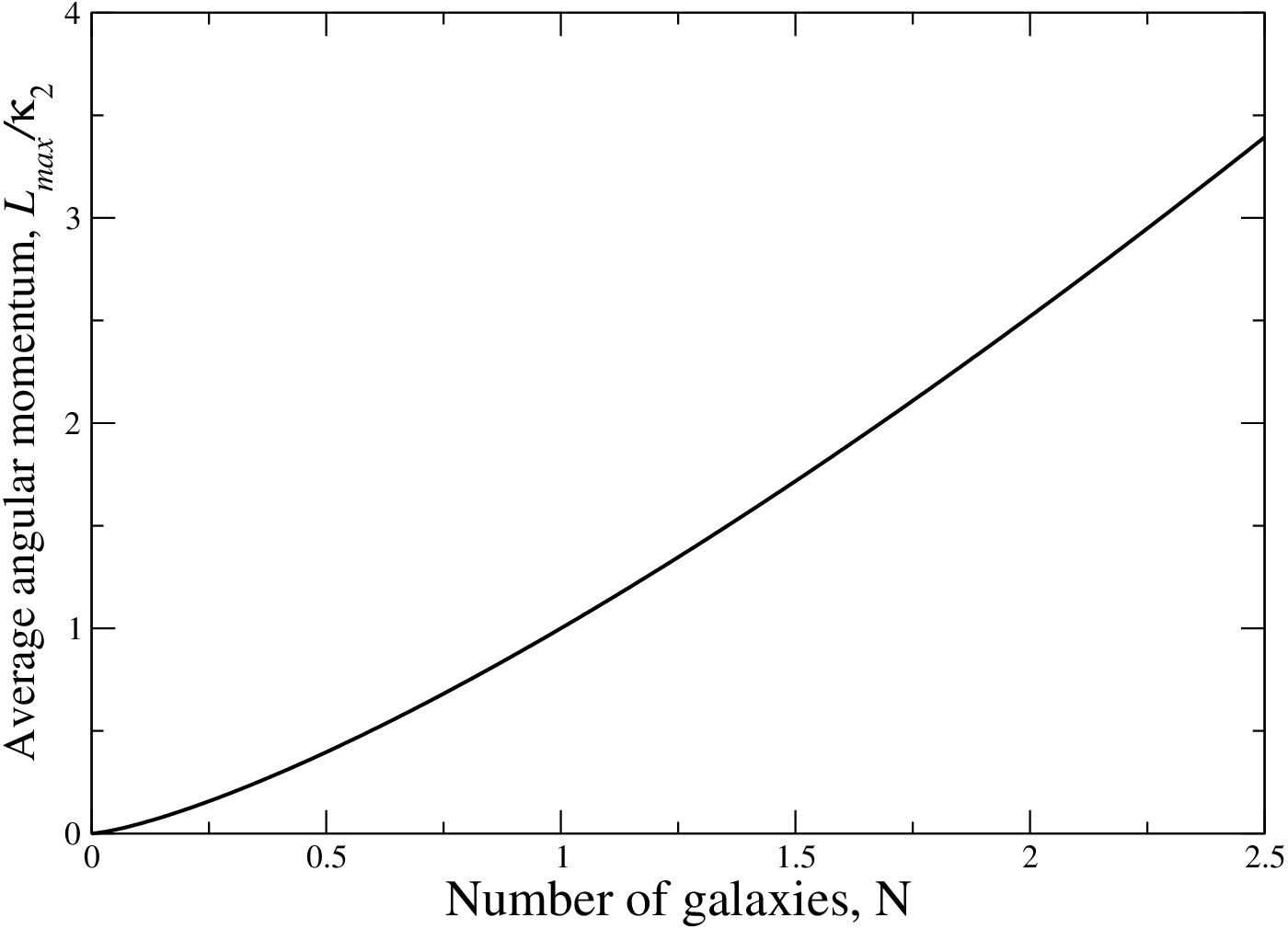}
\caption{Dependence of average galaxy angular moment $L_{max}=\kappa_2N^{4/3}$ on number of galaxies $N$ in a 
stellar system. $\kappa_2=$ $\kappa \frac{t}{t_0^2}GR^4m^2/V^{4/3}$.} \label{fig4}
\end{figure}

The equation \eqref{lm2} shows that mean orbital moment of a galaxy is proportional to $N^{4/3}$ where 
$N$ is number of galaxies. The dependence $L_{max}(N)=$ $\kappa_2N^{4/3}$ 
($\kappa_2=\kappa tGR^4m^2/(t_0^2V^{4/3})$) \eqref{lm3} is shown in Fig.\ref{fig4}. It is seen, that 
in our model of constant galaxies density $n=const$, the systems with larger number of galaxies have 
larger angular momenta. Below, comparing the linear and nonlinear regimes of fluctuation growth, we will 
show that the assumption of constant density may safely be used for qualitative analysis of angular momentum 
acquisition.
 
Let us finally pay attention that the dependence between mean momentum of galaxy cluster and its mass comes
 from parametrical dependence of $L_{max}$ on galaxy mass, volume and quadrupolar moment. This same dependence 
can be identically rewritten in different ways. Then, if we assume that certain 
parameters are constant, we obtain different dependences of $L_{max}$ on not only cluster mass $M$ but on mass 
and galaxies density as well as on galaxies number $N$. For instance, in Eq. \eqref{lm2}, at $R=$ const $L_{max}$ 
scales as a square of galaxy mass $m$ in contrast to relation \eqref{lm3}. This shows that observational verification 
of the dependences $L_{max}(N,m)$ would permit to make unambiguous conclusion about constancy of a particular stellar 
parameter. 

\subsection{Time dependence of distribution function in $\Lambda$CDM model.}

The time evolution of distribution function \eqref{star27b} can be obtained with respect to the definition of 
$\lambda$ \eqref{star26a} and subsequently $L_0$ \eqref{star7}. The dependence $L_0(t)$ generates  substitution 
$\lambda \to$ $\lambda/f_i(\tau)$, where $\tau=t/t_0$ so that we have from \eqref{star27b}
\begin{equation}\label{lcdm1}
H(\lambda,\tau)=\frac{2}{\pi \lambda}\int_0^\infty x\ \sin x \ \exp{\left[-\left(\frac {xf_i(\tau)}{\lambda}\right)^{3/4}\right]}dx, \ i=1,2.
\end{equation}
To derive Eq. \eqref{lcdm1} we take into account that there is additional coefficient $L_0$ in the denominator 
of \eqref{star27a} containing $f_i(\tau)$ so that there is no $f_i(\tau)$ before the integral in Eq. \eqref{lcdm1}.  
To obtain $f_{1,2}(\tau)$ in $\Lambda$CDM model, we begin with the determination of $a(t)$ from Friedmann 
equation \eqref{frid}, which reads
\begin{equation}\label{frid2}
\frac{da}{dt}=H_0\sqrt{\Omega_\Lambda a^2+\frac{1-\Omega_\Lambda}{a}}.
\end{equation}
The solution of the equation \eqref{frid2} has the form
\begin{equation}\label{frid3}
a(t)=\alpha \sinh^{2/3}(t/t_0),\ \alpha=\left(\frac{1-\Omega_\Lambda}{\Omega_\Lambda}\right)^{1/3},\
t_0=\frac{2}{3H_0\sqrt{\Omega_\Lambda}}.
\end{equation}

Now, the function $f_1(\tau)=$ $a^2(\tau)D'(\tau)$ ($D'=dD/d\tau$), where $D(\tau)$ can be found numerically 
from the equation \eqref{first}. Accordingly, in the nonlinear regime, the function $f_2(\tau)=E'(\tau)$ 
should be found numerically from the equation \eqref{sec}.

We note here, that in Einstein - de Sitter model $f_1(\tau)=(2/3)\tau$ and $f_2(\tau)=(-4/7)\tau^{1/3}$ 
\citep{dor70,ca96}. Also, the maximum of the function \eqref{lcdm1} occurs at 
\begin{equation}\label{lcdm2}
\lambda_{max}(\tau)\approx 0.602730263 f_i(\tau),
\end{equation}
where $\lambda_{max}\approx 0.602730263$ is a maximum of time - independent function $H(\lambda)$ \eqref{star28}.
It is seen that functions $f_2(\tau)$, related to the second perturbative corrections, are negative. 
Substitution of such function into the exponent of integrand \eqref{lcdm1} generates the imaginary part, 
which does not change the behaviour of $H(\lambda,\tau)$ qualitatively. That is why for the purpose of 
comparison of the linear and nonlinear regimes of fluctuation growth, everywhere we use the moduli of the 
functions $f_2(\tau)$ both in CDM and in $\Lambda$CDM models.
 
\begin{figure}
\centering
\includegraphics[width=0.9\columnwidth]{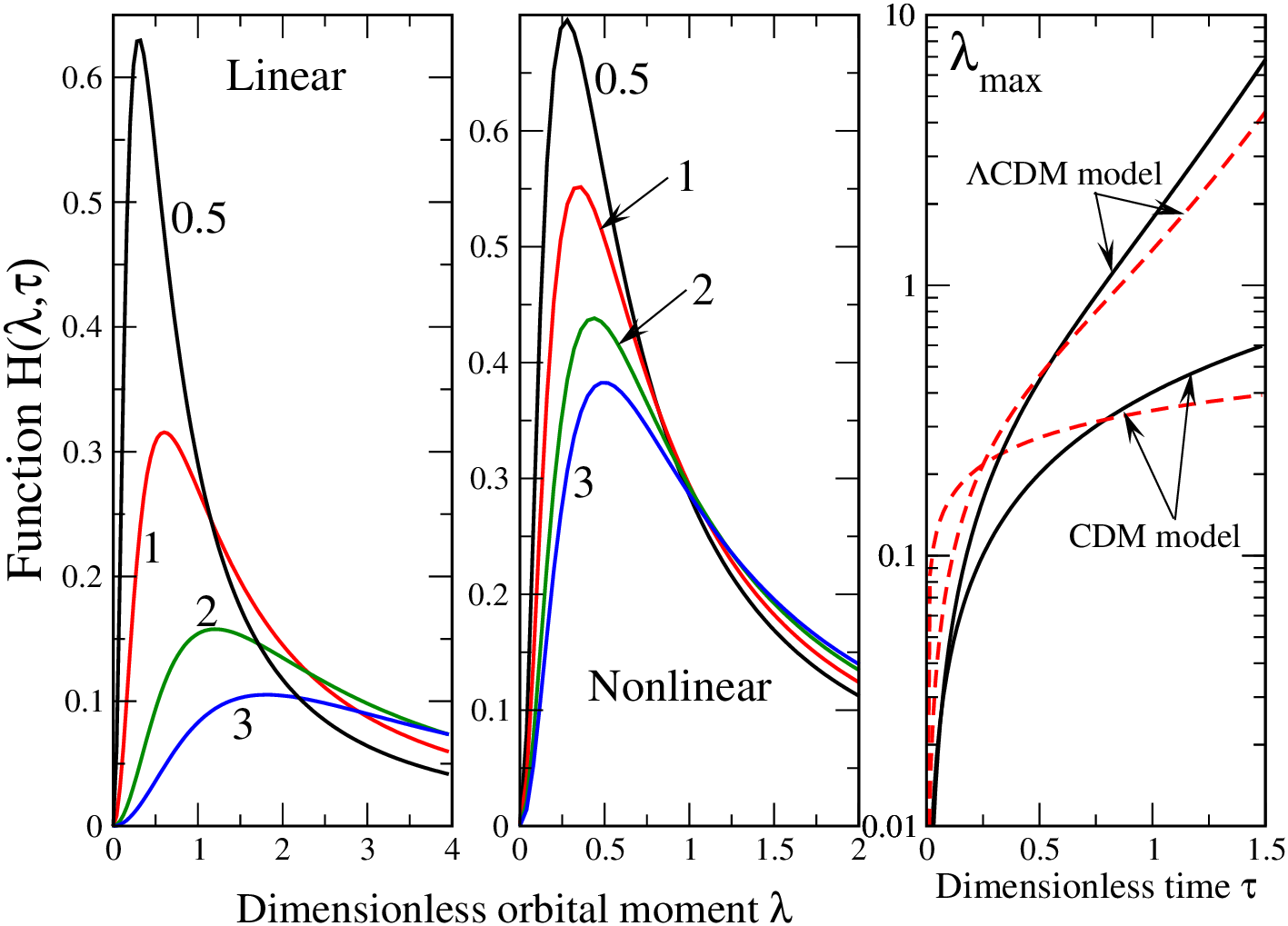}
\caption{Time evolution of effective 1D distribution function $H(\lambda,\tau)$ in CDM model. 
Left panel - linear regime $\lambda_{max}=0.6027f_1(\tau)$=0.4018$\tau$, middle panel - nonlinear regime 
$\lambda_{max}=0.6027$ $|f_2(\tau)|$ $=0.3444x^{1/3}$ (as $f_2(\tau)<0$, everywhere here we take its modulus). 
Figures near curves correspond to dimensionless time $\tau=t/t_0$. Right panel reports the dependence 
$\lambda_{max}(\tau)$ for CDM and $\Lambda$CDM models. Solid lines correspond to linear regime, 
dashed - to nonlinear. Mind the logarithmic scale on the vertical axis, needed to place the curves of CDM 
and $\Lambda$CDM models having very different growth rates. For $\Lambda$CDM model we choose 
$\alpha=$ $\left(\frac{1-\Omega_\Lambda}{\Omega_\Lambda}\right)^{1/3}=1$.} \label{fig2}
\end{figure}

The dependences $H(\lambda,\tau)$ \eqref{lcdm1} for Einstein - de Sitter CDM model with above 
$f_i(\tau)$ are shown in the Fig. \ref{fig2}. It is seen that at time growth the distribution function 
diminishes, while at zero time it goes to infinity. As time (figures near curves in Fig. \ref{fig2}) grows, 
the maximum $\lambda_{max}(\tau)$ shifts towards larger $\lambda$ so that the whole distribution function
 "blurs" at large times. This is related to the fact that functions $f_i(\tau)$ enter the exponent in the 
integrand \eqref{lcdm1}. The comparison of left and middle panels of Fig. \ref{fig2} show that the behaviour 
of $H(\lambda,\tau)$ is qualitatively similar in linear and nonlinear regimes of fluctuation growth. 
This means that for qualitative analysis we may safely use the linear regime. To further demonstrate that, 
at the right panel of Fig. \ref{fig2}, we report the dependence  $\lambda_{max}(\tau)$ \eqref{lcdm2}.
 It is seen that both in linear and nonlinear regime this function grows with time, although in
 $\Lambda$CDM model this growth is much faster so that to show both CDM and $\Lambda$CDM curves in 
one panel, we use the logarithmic scale.

\begin{figure}
\begin{center}
\includegraphics[width=0.8\columnwidth]{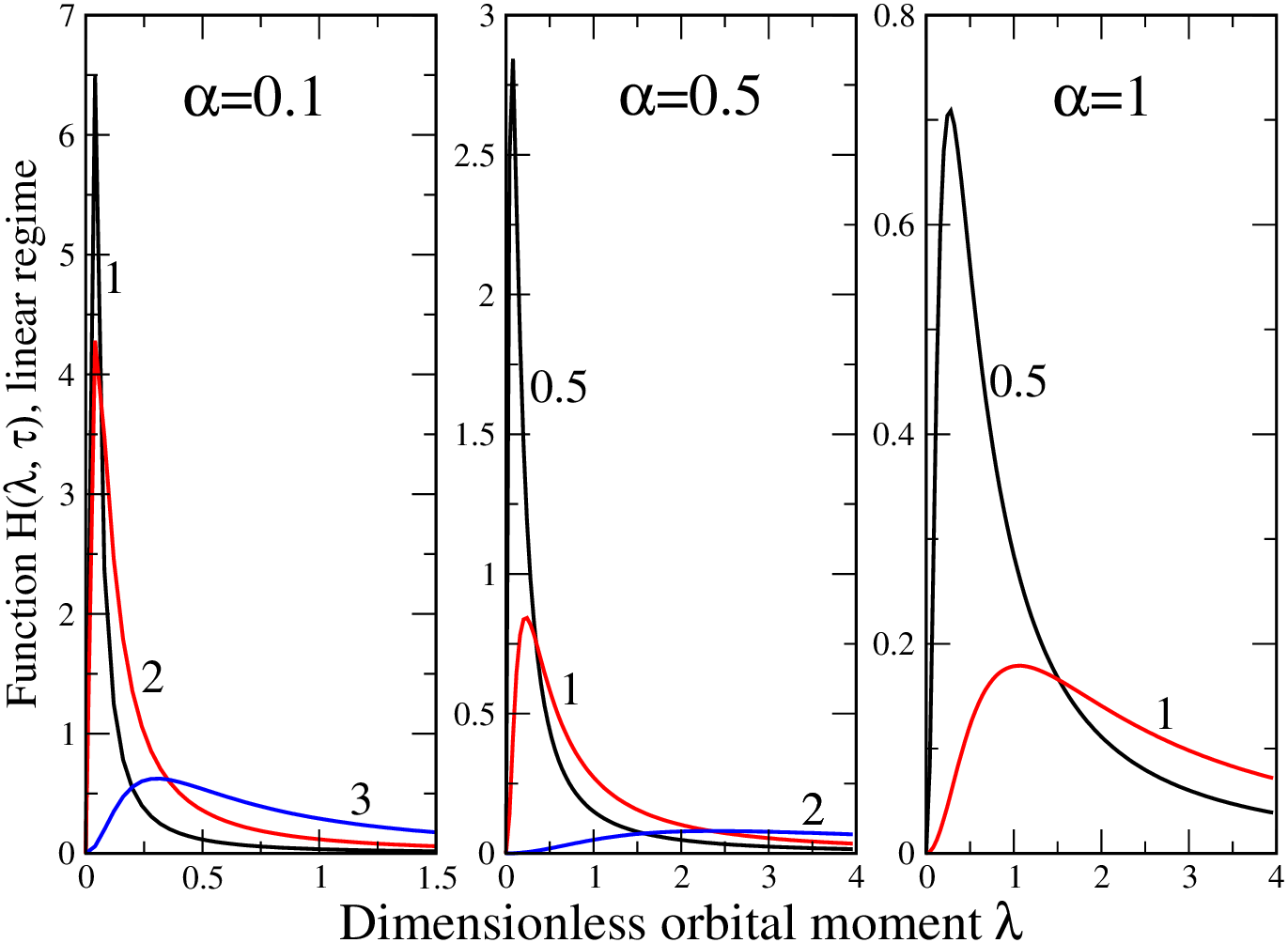} 
\end{center}
\vspace{5mm}
\begin{center}
\includegraphics[width=0.8\columnwidth]{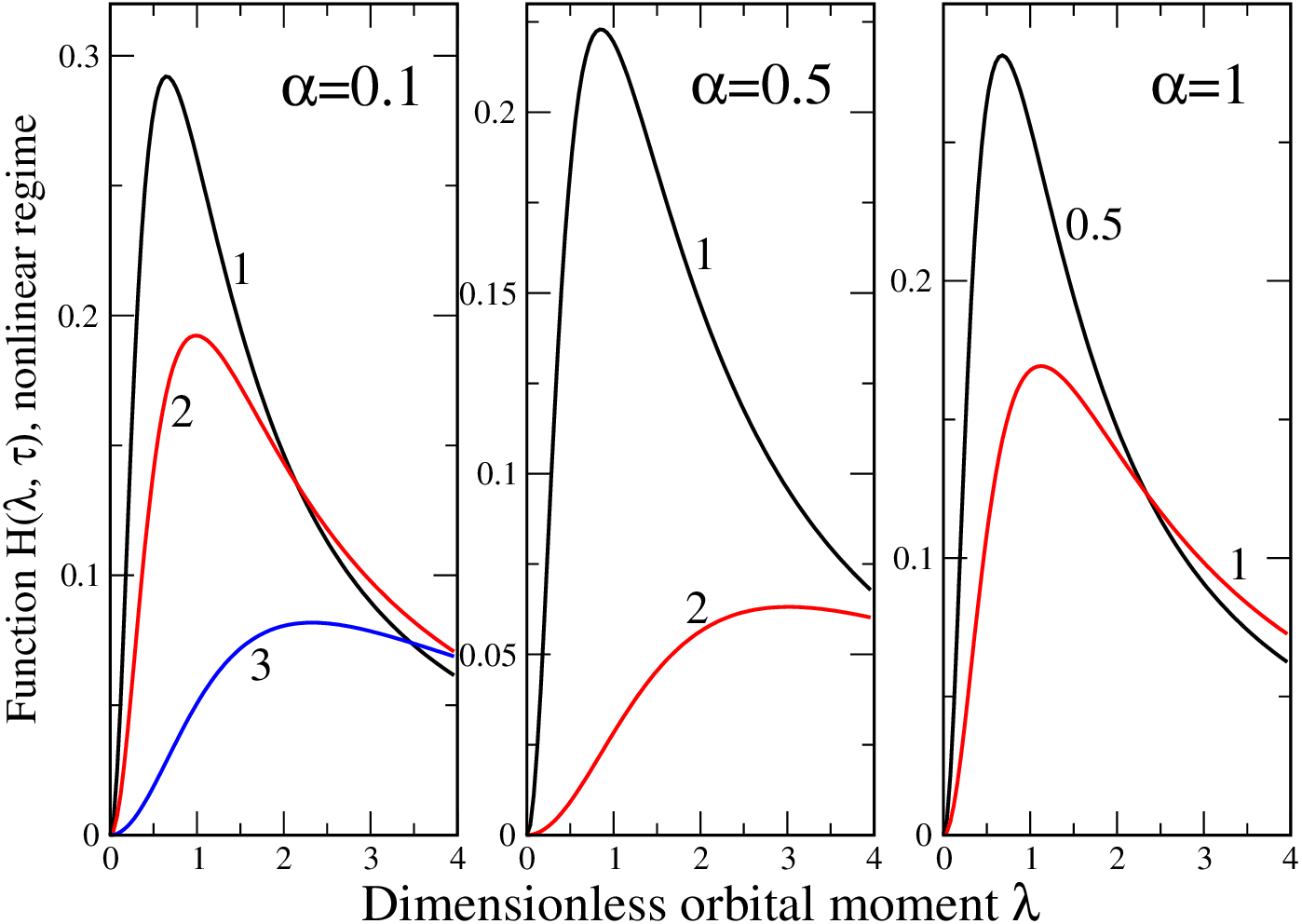}
\end{center}
\caption{Time evolution of effective 1D distribution function $H(\lambda,\tau)$ in $\Lambda$CDM model for 
different $\alpha$. Upper panels - linear regime, lower panels - nonlinear regime. The notations are the same 
as in Fig. \ref{fig2}.} \label{fig3}
\end{figure}

The result of numerical calculations in $\Lambda$CDM model are reported in Fig. \ref{fig3}. It is 
seen that qualitative behavior of $H(\lambda,\tau)$ is similar to that for CDM model. Here, however, 
we can trace the variation of $H(\lambda,\tau)$ with parameter $\alpha$. It is seen that as parameter 
$\alpha$ increases, the distribution function decreases - for the similar times the distribution function 
is smaller for larger $\alpha$. Similar to CDM model, the maximum of distribution function shifts towards 
larger $\lambda$ at time growth.  The above regularities are the same for linear and nonlinear regimes, 
while the values of distribution functions in $\Lambda$CDM model for the same $\alpha$ are smaller in 
nonlinear regime. This is related to the fact that function $f_2(\tau)$ grows much faster then 
$f_1(\tau)$ in $\Lambda$CDM model.
Also, as has been noticed in many references (see, e.g. \citet{cat96} and references therein), that first
 order perturbation result (linear regime) corresponds to so-called Zeldovich approximation which is 
approximately valid also for nonlinear situation. This shows once more that for qualitative discussion 
of the time dependence of the distribution function $H(\lambda, \tau)$ we can safely use the linear 
approximation, based on the function $f_1(\tau)$ only. It should be noted that parameter $t_0$ in Eq.\eqref{frid3}
 is of order of the Hubble time $1/H_0$. The comparison of the curves for different $\tau=t/t_0$ in the
 Fig. \ref{fig3} shows that the relaxation time is very long. This is in agreement with the fact  that 
clusters are known to be dynamically young objects, i.e. with crossing timescales which are non negligible 
with respect to the age of the Universe at the time of their formation. The fact that the relaxation time 
in Fig. \ref{fig3} is very long suggests that the dependence of the effective 1D distribution function 
$H(\lambda,\tau)$ (and average angular momentum as the result) on the redshift should be weak. Also, the 
comparison of Figs \ref{fig2} and \ref{fig3} shows that the distribution of spin parameters of galaxy 
clusters depends on the cosmological parameters and generally speaking is weak.

\section{Discussion. Relation to observational results}

Our theoretical results about the distribution function of momenta can shed some light on the problem 
of galaxies orientation in clusters. Our main message is that although the gravitational interaction between 
galaxies is of long-range character, the observations (which we will discuss below) may evidence that there 
is additional short-range intergalaxy interaction with characteristic radius $r_c$.  This means that if the 
distance $r$ between two galaxies is smaller then $r_c$, they are correlated and have their orbital moments 
aligned. This is the case for the dense (rich) galaxy clusters, which, by this virtue, have high degree of 
orbital moments alignment. In the opposite situation of poor clusters, where the intergalaxy distance $r>r_c$, 
the long-range multipole interaction of alternating sign dominates and the 
alignment of the orbital moments is absent. We speculate that this situation resembles that 
in diluted magnetic systems, where the presence of short-range exchange interaction between magnetic spins promote 
long-range magnetic order, which is characterized by macroscopic spin alignment, see \citet{stef97, semstef03} and 
references therein for details.  

The aforementioned statistical method permits to account for this situation if we add the (empirical) short-range 
interaction term to the initial potential \eqref{star3}. In this case, the distribution function of random fields 
would depend on the above average value of angular momentum $L_{max}\equiv L_{av}$ as a parameter (see \citet{stef97, semstef03}) 
so that self-consistent equation for $L_{av}$ of the form
\begin{equation}\label{lmax}
L_{av}=\int L(E) f(E,L_{av})d^3E
\end{equation}
can be derived. Here $f(E,L_{av})$ is the distribution function of random gravitational fields, which substitutes the 
expression \eqref{star19} in the case of inclusion of the possible short-range interaction term. In such case, for 
finite $r_c$, the distribution function decays at $E \to \infty$ faster then \eqref{star19} so that the integral \eqref{lmax} 
converges. As now the total interaction potential includes both luminous and dark matter, the equation \eqref{lmax} permits 
to address the question about alignment of sub-dominant galaxies, when most of cluster angular momentum is in the 
smooth dark matter halo component. Observationally it is relatively easy to analyse the orientation of 
angular momenta in the luminous matter i.e. in real galaxies and their clusters. With dark matter (sub) halos this 
is not so easy. One should not, however, forget, that there are relation between the properties of luminous and dark matter.
The articles  \citep{Paz08, Bett10, Kim11,Varela12} show clear observational evidence of 
relation between the dark matter halos and galaxies orientation,  see however \citet{Tenneti15} for opposite opinion. 
The assumption that angular momentum of the luminous matter traces that of the associated dark matter (sub)-structures 
(e.g. angular momentum totally aligned, one galaxy per sub-halo) allows us to conclude that angular momentum alignment 
of galaxies give us information about similar alignment in dark matter. In other words, the properties of angular momentum 
of luminous matter (like real galaxies) give us information about those of dark matter (sub)-structures.

For luminous matter, it is possible to consider the relation between the angular momentum and the mass of a structure as 
based on the observational data \citep{Godlowski05,g2011a}. It is possible to investigate how this image varies with the 
mass of galaxy clusters, beginning with the simplest ones, i.e. galaxies pairs. These investigations show that their angular
momentum originates mainly from the orbital motion of galaxies \citep{Kar84a,Kar84b,Min87}. \citet{Helou84} examining a 
sample of 31 galaxy pairs, found that an "anti alignment" of these galaxies spins occurs. \citet{Par97}
 recognized a weak alignment in physical pairs of galaxies. Alignment in pairs of spiral galaxies was also discerned by
 \citet{Pest04}. Intrinsic spin alignment in galaxies pairs has  independently been confirmed
 by \citet{Hey04} within their research on weak gravitational lensing, where it was necessary to estimate and
 remove the effects related to alignment of galaxy orientations. Also the analysis of positions of the Milky Way's companions
 shows their non random distribution (they are located perpendicularly to the Milky Way's disc), which can be regarded as 
their orbital alignment. Galaxies within compact groups rotate on prolate orbits about the group's centre 
\citep{Tovmas01}, which contributes to the system's total angular momentum. \citep{Yang06} found, while \citet{Sales09,Wang09,Wang10} 
confirmed it, that the companions of central red galaxies are aligned along their large axes.
The similar result has been obtained by \citet{Ibata13,Ibata14} in the two papers about the ordering of satellites orbits around M31 
Theirs latest article \citep{Ibata15} also corroborates this result. 
Thus it can be maintained that structures like galaxies and their companions, pairs and compact groups of galaxies
have a non zero net angular momentum related mostly to their orbital motion.
One should not forget, however, that situation is more complicated in more massive structures. 
As there is no evidence for rotation of the groups and clusters of galaxies (see, e.g., \citet{regos98,dia97,dia99,rines03,Hwang07}),
it is clear that angular momentum of such structures is related primarily to the alignment of constituents spins.

We should note here that there are many (seemingly contradictory) observational results regarding the alignment (or misalignment)
 of galaxies angular momenta in the literature beginning with the paper of \citet{t76}, who found an alignment in the galaxy 
orientations in the Virgo and A2197 clusters. The evidence for alignment of galaxies belonging to the Virgo cluster had 
been found by \citet{Helou82,mg85a}. Non random galaxy orientation was found in very rich galaxy clusters 
\citep{Djorgovski83,Godlowski98,Wu98,Baier03,Kit03,Aryal07,g2011b}. The alignment of galaxy planes was also found in A1689 
\citep{Hung10,Hung12}. This result is important as A1689 is the most distant cluster where the alignment has been found till now.

There are also contradictory results. For instance,  \citet{b88,b03,Hofman89} studied orientation of galaxies within 
clusters and did not find any alignment. The same result has been obtained by \citet{Aryal05c} in theinvestigations
 of three Abell clusters of richness class zero. During studies of the isolatedAbell groups \citep{Flin91,Trevese92,Kim01,NO10},
 only a rudimentary alignment was found and related only to the brightest cluster members. The alignment has 
also not been found during analysis of Tully's groups of galaxies belonging to the Local Supercluster \citep{g5,Godlowski05,g11c}. 
Summarizing above observations of angular momenta misalignment, we can conclude that such misalignment had been obtained 
for less massive structures like group and poor galaxy clusters.

To check the hypothesis, that the alignment of galaxies angular momenta increases with the cluster richness, \citet{g10a} 
 examined orientation  of galaxies  in  clusters both qualitatively and quantitatively. The analysis of the spatial 
orientations of  galaxies in the 247 optically selected rich 
Abell clusters, having in the considered area at least 100 members has been performed by \citet{g10a}. 
The structures have been taken from the PF catalog \citep{Panko06}. The statistical analysis, based on linear regression, 
permitted to conclude that cluster angular momenta increase with their numerousness.
Note however, that relatively small statistical sample of 247 clusters, analyzed by \citet{g10a}, does not give a possibility 
to discriminate between linear dependence tested by \citet{g10a}, the dependence $L_{max} \sim M^{5/3}$ (\citet{cat96})  
and our result \eqref{lm3} that mean angular momentum is proportional to $M^{4/3}$ (stemming from $N^{4/3}$). However,
 such detailed analysis would be possible if larger statistical sample of galaxy clusters is available.

The above results show clearly that galaxy clusters have a non zero net angular momentum related mostly to their orbital 
motion. For more massive structures there is a lack of 
alignment of the orientation of galaxies for group and poor galaxy clusters, while there is evidence
for alignment for the rich clusters of galaxies (\citet{Godlowski05}, see also \citet{g2011a} for 
later improved analysis). It has been suggested that degree of alignment increase with clusters richness, 
and as a result the cluster angular momentum increases with its numerousness. Here we emphasize once more that the 
equation \eqref{lmax}, which depends parametrically on the clusters parameters (like mass density $\rho$ and cluster 
density $n$) will give zero or nonzero solutions for $L_{av}$ (corresponding to alignment or misalignment of galaxies 
angular momenta) depending on the parameters values and cluster reachness in particular. 

Now we present an alternative explanation of alignment and misalignment of galaxies 
angular momenta in rich and poor clusters respectively. Namely, as rich galaxy clusters are less dynamically evolved 
objects, the galaxies constituting them, haven't had time to reach the pericenter of their orbit in the cluster and 
retain the alignment imprinted by the larger scale environment in which the cluster is embedded. In other words, 
large scale filaments (cosmic web) feed (reach) galaxy clusters along privileged directions, resulting in clusters 
being rather more prolate in shape than spherical.  We begin with the paper of \citet{Godlowski98}, who had shown 
that the galaxies orientation distribution in the Abell 754 double cluster is nonrandom with galaxy planes being 
perpendicular to the main cluster plane. The above nonrandom orientation distribution had also been found in the 
Abell 754 cluster \citep{Baier03}, but the direction (relatively to the main cluster plane) of the observed galaxy 
ordering is perpendicular to that for Abell 754. The interpretation of above orientation difference has been 
presented by \citet{diF88} in terms of two different types of galaxy clusters: oblate and prolate. One more 
interpretation can be done on the base of tidal interaction scenario. Namely, it has been observed by \citet{Paz08} 
that in large scale structure the direction of angular momenta (relatively to its main plane) of constituting 
objects depends on the structure mass. The same result has been obtained by \citet{Trujillio06, Varela12}. 
The newest analysis \citep{Codis12} (based on the earlier studies of \citet{sugerman00,Lee00,Bailin05,a1,Hahn07,Paz08,Zhang09}) 
on the dark matter halos angular momentum orientation also confirm the above result. The studies of galaxies 
angular momenta ordering in large scale had been fulfilled by \citet{Paz08,Zhang13}, who use the data from 
Sloan Digital Sky Survey catalog. They found that galaxies angular momenta align perpendicularly to the 
large scale structures planes. Latter effect has not been observed for the structures with relatively 
small masses. These results agree well with the simulations of \citet{Paz08} based on tidal interaction 
mechanism. \citet{Jones10} have found that the spins of spiral galaxies in the cosmic web have tendency 
to align along the filaments axes, which has been interpreted as the "fossil" evidence of the effects of 
long-range tidal interactions.
  
The other interesting problem is possible time evolution of galaxy clusters alignment. Assuming  
Einstein - De Sitter model \citep{dor70}, on the Fig. \ref{fig2} we report the time evolution of the function 
$H(\lambda)$.   It is interesting that at time growth the distribution function goes to zero. This means that 
older structures (clusters) should have more scattered observed values of angular momenta than younger ones. 
At the same time, equation \eqref{lm3} shows that mean angular momentum of the clusters should increase in time.  

Latter result is obtained in Einstein - De Sitter model but is  still valid for any similar cosmological models.
The predictions will be available to verify when we get better data concerning alignment in galaxy clusters. 
Note however that even now we have observational results suggesting the possible evolution of alignment with 
redshift. There are  for example  the results of \citet{song12} who found that the alignment profile of 
cluster galaxies drops faster at higher redshifts.

\section{Conclusions}

The main physical result of the present paper is the calculation of the distribution function of 
the gravitational fields of astronomical objects like galaxies ensembles and smooth halos based solely on the 
tidal interaction between constituting elements.  We show 
that for tidal (quadrupolar) interaction the distribution function cannot be Gaussian, its explicit form 
being presented by the equation \eqref{star19}. We emphasize here that derived distribution function 
of gravitational fields does not depend on the specific Eulerian or Lagrangian description of Newtonian matter 
and thus can be used to calculate virtually any observable characteristic of stellar ensemble. As an example, 
we use the above distribution function to calculate the distribution of angular momenta.
From the astronomical interpretation point of view, it is important that for particular cluster with richness 
$n$  we expect not a single value of angular momentum but the range of allowed values described
by the probability function. As the distribution function \eqref{star19} 
slowly decays at infinity, its first moment does not exist because the corresponding integral is divergent. 
To calculate the mean value of angular momentum in this situation, we assume that the maximal value of 
distribution function $L_{max}$ gives the desired quantity. We note that such procedure is usual while 
dealing with so-called heavy-tailed distributions, see, e.g. \citet{kapkes92} and references therein. 

As astronomical objects (for instance galaxies) masses, radii and number enter the equation 
\eqref{star19} as parameters, we were able to show 
that mean value of angular momentum $L_{max}$ for particular galaxy
cluster of mass $M$ is proportional to $M^{5/3}$, thus corroborating well-known (see \citet{Sch09} and 
references therein) empirical result. Our other result $L_{max} \sim $ $N^{4/3}$ (or $M^{4/3}$, eq. \eqref{lm3})
 has also its astronomical interpretation that larger (richer) clusters of galaxies have higher angular momentum, 
see also the discussion below. The observational discrimination between the dependences $M^{5/3}$ and $M^{4/3}$ 
would be possible when larger statistical sample of galaxy clusters will be available. The parametric time 
dependence of $L_{max}$ via functions $a(t)$ and $D(t)$ in Einstein - de Sitter model permits us to trace its
 time evolution.  This shows that our approach to derivation of distribution function of galaxies 
angular momenta gives physically reasonable answers. We have also analyzed the time evolution of the distribution 
function. It is reported in Fig. \ref{fig2} and shows that the distribution function is flattening with time.

The relation between angular momentum and mass of the structures has been extensively analysed theoretically 
\citep{Muradyan75,Wesson79,Wesson81,Wesson83,Carrasco82,Sistero83,Brosche86,Mackrossan87,Paz08}, 
see \citet{Sch09} for review. This relation has usually been presented empirically in the form $L_{max} \sim M^{5/3}$. 
From the point of view of modern scenarios of galaxy and their structures formation,  increasing of angular 
momentum with the cluster richness could be explained only in tidal torque scenario in the hierarchic clustering
model \citep{heav88,Hwang07,Noh06a,Noh06b}  and in  Li model \citep{Li98,Godlowski05}. One should note 
however that the value of the Universe rotation, required by  \citet{Li98},  is too large
as compared to the anisotropy found in cosmic microwave background radiation (CMBR).

The increase of the galaxies angular momentum with mass of the structure was found observationally  
during analysis of the alignment of galaxies in clusters \citep{Godlowski05,Aryal07,g10a,g2011a,g11c,g2011b}. 
Since it is commonly agreed that groups and clusters of galaxies do not rotate  
\citep{regos98,dia97,dia99,rines03,Hwang07}, any possible nonzero angular momentum of such structures should arise 
only from possible alignment of galaxy spins and stronger alignment mean greater angular momentum of 
such structures. Generally there is no evidence for a non zero angular momentum of galaxies or group and poor
 galaxy clusters, while we observe such alignment  for rich clusters of galaxies and  superclusters. We speculate
 that this phenomenon may occur due to some additional short-range interaction between galaxies such that in rich 
clusters the galaxies are correlated as they fall in the range of this interaction and hence have their 
angular moments aligned. In such situations there should be some critical richness $n_{cr}$, related to the 
interaction range $r_c$ such that only clusters of richness $n>n_{cr}$ would have their spins aligned. We note 
here that such physical picture is common for disordered magnets and ferroelectrics, see \citet{stef97, semstef03} 
and references therein. 
We postpone the quantitative investigation of this interesting question for future publications. 
The above scenario can we applied to the problem of the possible merger of galaxies into larger objects. 
The corresponding results will also be published elsewhere. 

The problem of galaxies merger in a cluster is related to that of the role of much more massive central galaxy. 
This problem, in turn, is due to the fact of (generally speaking random) interaction between cluster members 
and dynamic evolution of the nearby (to specific cluster member) structures. The problem of dynamic evolution
 can be studied by the combined examination of mutual orientation of galaxies in clusters and Binggeli effect  
\citep{b2,s2,c1,h3,g10}. More specifically, this can be done by two methods. The first is Binggeli effect 
studies, i.e. the investigation of relation between positions of great axes in groups or clusters of galaxies 
and directions towards their neighbors. The second one is the studies of mutual orientation of the brightest 
galaxy (and other bright galaxies) in a structure relatively to the position of cluster great axes or even 
examination of structure ellipticity redshift dependence, especially in the enlarged observational samples.
The analysis of the differences between position angles of the Tullys groups of galaxies belonging to the Local 
Supercluster shows \citep {g10} that there exists the alignment of the line joining two brightest galaxies with 
both the position angle of the parent group and the direction toward Virgo cluster center. This analysis reveals 
the following picture of the structure formation. Two brightest (most massive) galaxies were formed firstly. 
They originated in the filamentary structure directed toward the center of the protocluster. This is the place 
where the Virgo cluster center is located now. Due to gravitational clustering, the groups were formed in such 
a manner that galaxies follow the line determined by the two brightest (most massive) objects. Therefore, the 
alignment of the structure position angle and line joining two brightest galaxies is observed. The other groups 
were formed on the same or nearby filament. This shows the particular role played by the more massive (brighter) 
galaxies. In our future investigations, we will analyze the particular role played by the central (most massive) 
galaxies quantitatively.

One should note that the theoretical and observational analysis of galaxies alignment in cluster is also very 
important from the point of view of weak gravitational lensing \citep{Troxel14,Joachimi15}. There are mutual influence
of the orientation of galaxies and weak gravitational lensing. It should be pointed out as well that the examination 
of galaxies orientations is also meaningful due to one of the outcomes of the activity of gravitational lensing effect 
 \citep{heav00,Schn05} which is the alignment of the galaxies images. Such alignment, in several Mpc scale is also 
expected in case of cosmic shear existence. \citet{critt01}  proved that at least in the scenario of tidal 
interactions, the effects of alignment can be distinguished from the effects of weak gravitational
lensing. Taking both of these effects 
into consideration (in appropriate proportions) is of a paramount importance for mapping the mass distribution 
with weak lensing techniques, and vice versa: weak lensing induced shape deformations are important for studies 
of the intrinsic orientation of galaxies within structures. Therefore weak-lensing studies 
will allow investigating mass distribution in clusters which is important for studies of dark matter in them. 

Let us finally summarize the simplifying assumptions made in the above calculation of 
the distribution function. We assume that autocorrelation radius is constant and is the same
 for each cluster \citep{autocor,autocor1,astr,lin,tucker,long08}. Also we treated cluster density as a parameter 
rather then a function of interobject (intergalaxy) distance. We note here that as time (via the 
functions $f_{1,2}(t)$, calculated in CDM and $\Lambda$CDM models in linear and nonlinear fluctuation growth regimes) 
enters the distribution function as a parameter, our approach is valid for any cosmological model - from conventional
 Einstein - De Sitter CDM \citep{dor70} to $\Lambda$CDM. Most important simplification is to assume that all galaxy 
has equal masses. In the future studies we plan to consider more realistic situation, introducing $n=n({\bf r})$, the 
real (i.e. extracted from observational data) galaxy mass distribution function and the short-range term in the potential 
of interaction between astronomical objects (say galaxies and their clusters). Such generalisations will 
require numerical calculations of the distribution function of gravitational fields and its mean value. 

\section*{Acknowledgements}
We are grateful to anonymous referee, whose expertise helps us to improve our work substantially.

 \clearpage 

\clearpage

\clearpage

\clearpage

\end{document}